\documentclass[11pt,iop]{emulateapj}
\shorttitle{EEP constraints with TeV Blazars}
\shortauthors{Wei et al.}

\begin{document}

\title{Tests of the Einstein Equivalence Principle using TeV Blazars}
\author{Jun-Jie Wei\altaffilmark{1}, Jie-Shuang Wang\altaffilmark{2}, He Gao\altaffilmark{3}, and Xue-Feng Wu\altaffilmark{1,4}}

\affil{$^1$ Purple Mountain Observatory, Chinese Academy of Sciences, Nanjing 210008, China; jjwei@pmo.ac.cn, xfwu@pmo.ac.cn \\
$^2$ School of Astronomy and Space Science, Nanjing University, Nanjing 210093, China\\
$^3$ Department of Astronomy, Beijing Normal University, Beijing 100875, China\\
$^4$ Joint Center for Particle, Nuclear Physics and Cosmology, Nanjing
University-Purple Mountain Observatory, Nanjing 210008, China}

\begin{abstract}
The observed time delays between different energy bands from TeV blazars provide
a new interesting way of testing the Einstein Equivalence Principle (EEP). If
the whole time delay is assumed to be dominated by the gravitational field of the
Milky Way, the conservative upper limit on the EEP can be estimated. Here we show
that the strict limits on the differences of the parameterized post-Newtonian parameter
$\gamma$ values are $\gamma_{\rm TeV}-\gamma_{\rm keV}<3.86\times10^{-3}$ for
Mrk 421 and $\gamma_{\rm TeV}-\gamma_{\rm keV}<4.43\times10^{-3}$ for Mrk 501,
while expanding the scope of the tested EEP
energy range out to the TeV--keV range for the first time. With the small time lag
from the 0.2--0.8 TeV and $>0.8$ TeV light curves of PKS 2155-304, a much more severe
constraint on $\gamma$ differences of $\sim10^{-6}$ can be achieved, although the
energy difference is of order of $\sim$ TeV. Furthermore, we can combine these
limits on the energy dependence of $\gamma$ with the bound on the absolute $\gamma$
value $\gamma-1\sim0.3\%$ from light deflection measurements at the optical (eV)
bands, and conclude that this absolute bound on $\gamma$ can be extended from
optical to TeV energies.
\end{abstract}


\section{Introduction}

The Einstein Equivalence Principle (EEP), which is one of the major pillars of general relativity
and other metric theories of gravity, says that the trajectories of freely falling uncharged test bodies
are independent of their internal compositions and structures. The possible violations of EEP
would have significant impact on people's understanding of nature, it is therefore important to
sequentially improve the verification of its accuracy.

The validity of the EEP and general relativity in a post-Newtonian
context can be characterized by limits on the numerical coefficients of the parameterized post-Newtonian
(PPN) parameters, such as the parameter $\gamma$. Here $\gamma$ is defined as how much space curvature
produced by unit rest mass (see Will 2006, 2014, for a recent review).
More specifically, one can test the accuracy of EEP and general relativity by measuring the absolute value
of $\gamma$ (e.g., Froeschle et al. 1997; Bertotti et al. 2003; Lambert \& Le Poncin-Lafitte 2009, 2011),
as well as constraining the differences of $\gamma$ values for different types of massless
(or negligible rest mass) neutral particles, or for the same type of particle with different energies
(e.g., Krauss \& Tremaine 1988; Longo 1988; Gao et al. 2015; Wei et al. 2015), since general relativity
predicts $\gamma\equiv1$ and all gravity theories incorporating the EEP also predict
$\gamma_{1}=\gamma_{2}\equiv\gamma$, where the subscripts correspond to
two different test particles.

Determinations of the absolute $\gamma$ values have reached high precision from light deflection
and time delay measurements. The light deflection measurements from the very-long-baseline
radio interferometry yielded $\gamma-1=(-0.8\pm1.2)\times10^{-4}$ (Lambert \& Le Poncin-Lafitte 2009, 2011).
Through the time delay measurements of a radar signal from the Cassini spacecraft, Bertotti et al. (2003)
obtained an accurate determination of $\gamma-1=(2.1\pm2.3)\times10^{-5}$. These results are
in good agreement with the prediction of general relativity. \footnote{Note that some other
gravitational theories besides general relativity also predict $\gamma\equiv1$ (Will 1993).}
On the other hand, some astronomical sources have been used to test the EEP by comparing the
$\gamma$ values for different test particles in a few instance, including the following
representative cases: Krauss \& Tremaine (1988) and Longo (1988) proposed that the observed
time delay of the neutrinos and photons from supernova 1987A provides a new test of the EEP,
they presented an upper limit on $\gamma$ differences of 0.34\% for neutrinos and photons,
and a more severe limit of $1.6\times10^{-6}$ for neutrinos ranging in energy from 7.5 to 40 MeV;
Gao et al. (2015) used the time delays between correlated photons from gamma-ray bursts (GRBs)
to constrain the accuracy of the EEP and found that the differences of the $\gamma$ values for
photons over the MeV--GeV or eV--MeV range is as low as $\sim 10^{-7}$, improving the limits
from supernova 1987A by at least one order of magnitude; and Wei et al. (2015) proved that
fast radio bursts (FRBs) of extragalactic origin can serve as an ideal testbed to probe
the EEP and set the most stringent limit on $\gamma$ differences up to now, yielding $\sim 10^{-8}$,
by analyzing the arrival time delay of FRB photons of different frequencies,
which is at least 10 to 100 times tighter than the constraints from supernova 1987A and GRBs.

It is well known that blazars are an extreme subclass of active galactic nuclei,
which can be further divided into flat spectrum radio quasars if they have strong
emission lines and BL Lacertae objects if they have weak or no emission lines
(e.g., Ulrich et al. 1997).
Blazars are characterized by broadband non-thermal emission extending from
radio up to high-energy and very-high-energy (VHE) gamma-rays, and display of
violent variability on different timescales from minutes to years
(e.g., Wagner \& Witzel 1995). The broadband radiation is produced by
a relativistic jet pointed along the line of sight (e.g., Begelman et al. 1984;
Urry \& Padovani 1995). Because of their fast flux variability, cosmological distances,
and VHE photons in the TeV range, TeV blazars have been deemed as
an effective way to probe the effect of Lorentz invariance violation (LIV; e.g.,
Biller et al. 1999; Aharonian et al. 2008; MAGIC Collaboration et al. 2008; H.E.S.S.~Collaboration et al. 2011).
Here, we suggest that TeV blazars can also provide a good astrophysical laboratory
to constrain the EEP, which can further extend for the first time the scope of
the tested energy range out to TeV energies.

It is worth pointing out that testing the EEP with both GRBs and TeV blazars is of
great fundamental interest. GRBs can be detected out to very high redshifts ($z\sim8.2$),
but with very few high energy ($E>$ GeV) photons. On the contrary, TeV blazars can be well observed by
ground based detectors with large statistics of photons above a few tens of TeV.
But, since high energy photons would be absorbed by extragalactic background light,
TeV observations are limited to sources with low redshifts. Hence, GRBs and TeV blazars
are mutually complementary in constraining the EEP, and they enable us to test different redshift
and energy ranges. In this work, we first try to test the accuracy of the EEP using TeV blazars.
The rest of this paper is arranged as follows. In Section~2, we briefly describe the method
of testing the EEP. The constraints on the EEP from TeV blazars are showed in Section~3.
Finally, we summarize our conclusions in Section~4.

\section{Method description}

The observed time delays between different energy bands from the cosmological sources
have been used to set bounds on the EEP. In principle, the observed time delay
$(\Delta t_{\rm obs})$ (Gao et al. 2015; Wei et al. 2015)
\begin{equation}
\Delta t_{\rm obs}=\Delta t_{\rm int}+\Delta t_{\rm LIV}+\Delta t_{\rm spe}+\Delta t_{\rm DM}+\Delta t_{\rm gra}
\end{equation}
has contributions from the intrinsic time lag $(\Delta t_{\rm int})$, the LIV induced time delay
$(\Delta t_{\rm LIV})$, the possible time delay due to the non-zero mass of photons in special relativity
$(\Delta t_{\rm spe})$, the underlying time delay arised from the dispersion by the line-of-sight free electron content
$(\Delta t_{\rm DM})$, and the travel time delay $(\Delta t_{\rm gra})$ between energy bands $E_1$
and $E_2$, caused by an external gravitational potential $U(r)$, respectively. Among these terms,
\begin{equation}
\Delta t_{\rm gra}=\frac{\gamma_{\rm 1}-\gamma_{\rm 2}}{c^3}\int_{r_o}^{r_e}~U(r)dr
\end{equation}
is the relevant one to probe the EEP. Here, $\gamma$ is the PPN parameter, $r_{o}$ and $r_{e}$
represent locations of Earth and source. For high energy photons, such as the gamma-ray signals
considered here, $\Delta t_{\rm DM}$ is absolutely negligible. In addition, considering that both
$\Delta t_{\rm LIV}$ and $\Delta t_{\rm spe}$ are also negligible for the analysis of this work,
and assuming $\Delta t_{\rm int}>0$, one can derive
\begin{eqnarray}
\Delta t_{\rm obs}>\frac{\gamma_{\rm 1}-\gamma_{\rm 2}}{c^3}\int_{r_o}^{r_e}~U(r)dr\;.
\label{eq:deltatnew}
\end{eqnarray}
We refer the reader to Gao et al. (2015) for more details.

Generally, $U(r)$ should have three components, i.e., $U(r)=U_{\rm MW}(r)+U_{\rm IG}(r)+U_{\rm host}(r)$,
including the gravitational potentials of the Milky Way, intergalactic background, and host galaxy of the cosmological source,
respectively. Although the potential models of $U_{\rm IG}(r)$ and $U_{\rm host}(r)$ are hard to know,
it is plausible to assume that the effect of these two components is much greater than
if we just consider the potential of the Milky Way $U_{\rm MW}(r)$.
Adopting the Keplerian potential for the Milky Way, it would be reasonable to have
\begin{equation}
\gamma_{1}-\gamma_{2}<\Delta t_{\rm obs}\left(\frac{GM_{\rm MW}}{c^{3}}\right)^{-1}\ln^{-1}\left(\frac{d}{b}\right) \;,
\label{eq:gammadiff}
\end{equation}
where $M_{\rm MW}\simeq6\times10^{11}M_{\odot}$ is the total mass of the Milky Way (McMillan 2011; Kafle et al. 2012),
$b$ represents the impact parameter of the rays, and $d$ denotes the distance from the Earth to the source.

\section{Tests on the EEP from TeV Blazars}
As we know, mankind's view of nature would be greatly affected if the EEP is violated.
Thus, it is important to constantly test the validity of the EEP with all kinds of alternative
astronomical sources. We continue to search for such sources and propose that TeV blazars
are a new interesting tool for constraining the EEP, while extending the tested EEP energy range
out to TeV energies (more on this below). As examples, we use three famous TeV blazars
(Mrk 421, Mrk 501, and PKS 2155-304) to constrain the EEP accuracy.

\subsection{Mrk 421}
Markarian 421 (Mrk 421; $z=0.031$) is the first extragalactic TeV blazar to be detected at gamma-ray
energy $E>500$ GeV (Punch et al. 1992), with coordinates (J2000) $\rm R.A.=11^{h}04^{m}19^{s}$,
$\rm Dec.=38^{\circ}11^{'}41^{''}$. \footnote{The location information of TeV blazars are available
in the TeGeV Catalogue at http://www.asdc.asi.it/tgevcat/.}
Using the High Altitude GAmma Ray (HAGAR) telescope array, Shukla et al. (2012) observed
Mrk 421 in its high state of flux activity during February 13--19, 2010 and also detected
a very bright flare above 0.25 TeV. They investigated the correlation between light curves
of different energies with the cross-correlation function and found that VHE gamma-ray
($>0.25$ TeV) flux reaches peak with a $1.3$ days lag compared with the peak time of X-ray
(1.5--12 keV) flare.
With the measured time delay $\Delta t_{\rm obs}=1.3$ days and location information,
we thus obtain EEP constraint from Equation~(\ref{eq:gammadiff}) for Mrk 421
\begin{equation}
\gamma_{\rm TeV}-\gamma_{\rm keV}<3.86\times10^{-3}\;.
\end{equation}

\subsection{Mrk 501}
The TeV blazar Markarian 501 (Mrk 501) is a nearby, bright X-ray emitting source at $z=0.034$,
also well known to emit VHE ($E\geq100$ GeV) gamma-ray photons (Quinn et al. 1996),
with coordinates (J2000) $\rm R.A.=16^{h}53^{m}52.2^{s}$, $\rm Dec.=39^{\circ}45^{'}36^{''}$.
Furniss et al. (2015) reported on multiwavelength observational campaign of Mrk 501 between 2013
April 1 and August 10, including the first display of hard X-ray variability with $Swift$
and $NuSTAR$. The discrete correlation function was applied to study the cross-correlations between
different energy bands. A time lag of $0\pm1.5$ days was measured between the VHE observations
($>0.2$ TeV) and the 0.3--3 keV $Swift$/XRT band. To be conservative, we adopt the largest value
1.5 days as the time delay $\Delta t_{\rm obs}$ between these two energy bands.
With the above information of Mrk 501, a severe limit on the EEP from Equation~(\ref{eq:gammadiff}) is
\begin{equation}
\gamma_{\rm TeV}-\gamma_{\rm keV}<4.43\times10^{-3}\;.
\end{equation}

\subsection{PKS 2155-304}
The source PKS 2155-304 is one of the brightest TeV blazars and located at $z=0.117$,
almost four times more distant than Mrk 421 and Mrk 501. It was discovered
at the radio bands as part of the Parkes survey (Shimmins \& Bolton 1974), and
identified as a BL Lacertae object by Hewitt \& Burbidge (1980), with coordinates (J2000)
$\rm R.A.=21^{h}58^{m}52.6^{s}$, $\rm Dec.=-30^{\circ}13^{'}18^{''}$.
The observation of the VHE flare of PKS 2155-304 on July 28, 2006 by the High Energy
Stereoscopic System (H.E.S.S) provides the current best limit on LIV derived from
the observation of blazars (H.E.S.S.~Collaboration et al. 2011). Aharonian et al. (2008)
determined the time delay from two light curves in different energies using
the modified cross correlation function. In order to keep good photon statistics
in both two energy bands, while optimizing the energy gap between the two,
the correlation analysis was performed on the light curves between 0.2--0.8 TeV
and $>0.8$ TeV. A 95\% confidence limit on a linear energy dispersion of
73 s $\rm TeV^{-1}$ is thus given. Since the mean difference of the photon energies
between these two bands is 1.0 TeV, the dispersion per energy can be transformed to
the measured time delay $\Delta t_{\rm obs}=73$ s. From Equation~(\ref{eq:gammadiff}),
we can tighten the constraint on the EEP to
\begin{equation}
\left[\gamma(0.2\;\rm TeV-0.8\; \rm TeV)-\gamma(>0.8\; \rm TeV)\right]<2.18\times10^{-6}
\end{equation}
for PKS 2155-304, which is as good as the results on supernova 1987A from Longo (1988).

\section{Summary and discussion}

The accuracy of the EEP at the post-Newtonian level can be tested through the relative
differential variations of the PPN parameters, such as the parameter $\gamma$.
Inspired by the work of Gao et al. (2015), we continue to search for other alternative
astronomical sources that are suitable for testing the EEP accuracy. In this work,
we propose that TeV blazars can serve as a new good candidate for this purpose.
Furthermore, GRBs and TeV blazars are complementary to each other in constraining the EEP,
since they are observed in different energy and redshift ranges and with different
levels of variability.

Using the observed time delay $\Delta{t}\sim1.5$ days for the light curves with energy bands of
keV to TeV and a assumption that the time delay is dominated by the gravitational potential
of the Milky Way, we place robust limits on $\gamma$ differences for two TeV blazars, i.e.,
$\gamma_{\rm TeV}-\gamma_{\rm keV}<3.86\times10^{-3}$ for Mrk 421 and
$\gamma_{\rm TeV}-\gamma_{\rm keV}<4.43\times10^{-3}$ for Mrk 501.
On the basis of the time delay from two light curves between 0.2--0.8 TeV and $>0.8$ TeV bands,
we can tighten the limit on $\gamma$ differences to
$\left[\gamma(0.2\;\rm TeV-0.8\; \rm TeV)-\gamma(>0.8\; \rm TeV)\right]<2.18\times10^{-6}$
for PKS 2155-304, but the energy difference is of order of $\sim$ TeV.
It should be underlined that these upper limits are based on very conservative
estimates of the observed time delay and the total gravitational potential.
The inclusion of contributions from the neglected components in the observed
time delay (see Equation~1) could improve these limits in some degree.
In addition, if we have a better understanding of the host galaxy and
the intergalactic gravitational potential and these two effects are
taken into considered, our constraint results would be significantly
improved by orders of magnitude.

In the past, tests of the EEP through the differences of the $\gamma$ parameter
have been made of emissions from supernova 1987A (Krauss \& Tremaine 1988; Longo 1988),
GRBs (Sivaram 1999; Gao et al. 2015), and FRBs (Wei et al. 2015). In particular,
the observation of FRBs not only provided the most stringent limits on the accuracy
of the EEP, showing that the EEP is obeyed to the level of $\sim10^{-8}$, but also
extended the EEP tested energy range down to the radio band (Wei et al. 2015).
Also, Gao et al. (2015) extended the tested energy range up to the MeV--GeV and eV--MeV range
with the help of high energy photons from GRBs, although the capability of GRBs for
testing the EEP $(\sim10^{-7})$ was not so strong as that of FRBs. In the present paper,
we prove that TeV blazars are a new support tool for probing the EEP in different energy
and redshift ranges, and we derive the upper limits of $\sim10^{-3}$ for light curves
over the TeV--keV range and of $\sim10^{-6}$ for light curves over an energy range
between 0.2--0.8 TeV and above 0.8 TeV. Although our constraints on the EEP accuracy
are not as tight as previous results of GRBs or FRBs, the tested energy range can be
further extended out to the TeV--keV range by using the measured time delays of TeV blazars.

As mentioned above, the most precise limits on the absolute $\gamma$ value of photons
yield an agreement with general relativity of $\gamma-1\sim10^{-5}$, which is from radio photons
(Bertotti et al. 2003). And the light deflection measurements using the Hipparcos optical
astrometry satellite have reached $\gamma_{\rm eV}-1\sim0.3\%$ (Froeschle et al. 1997).
On the other hand, by constraining the differences of $\gamma$ values between various
energies, Gao et al. (2015) found that the absolute $\gamma$ values of MeV or GeV photons
differ from that of optical (eV) photons by $<10^{-7}$, and our results show that
the absolute $\gamma$ values of TeV photons differ from that of keV photons by $<10^{-3}$.
Combining these results, we can predict that the limits on the absolute $\gamma$ values of
TeV photons should also be consistent with general relativity to the same level of
$\sim0.3\%$, i.e.,  $\gamma_{\rm TeV}-1\sim0.3\%$. We therefore conclude that
this absolute bound on $\gamma$ can be extended from the optical to the TeV range,
and the value of $\gamma$ is identical for photons between optical and TeV to within
approximately $10^{-3}$.

\acknowledgments
We are grateful to the anonymous referee for his/her useful suggestions,
which have led to an improvement in the presentation of our manuscript.
We also acknowledge Houdun Zeng for helpful communications.
This work is partially supported by the National Basic Research Program (``973'' Program)
of China (Grants 2014CB845800 and 2013CB834900), the National Natural Science Foundation
of China (grants Nos. 11322328, 11433009, and 11543005),
the Youth Innovation Promotion Association (2011231), and the Strategic Priority Research Program
``The Emergence of Cosmological Structures'' (Grant No. XDB09000000) of
the Chinese Academy of Sciences.

\end{document}